\documentclass[12pt,preprint]{aastex}
\usepackage{natbib,psfig,emulateapj5}

\newcommand{\gtsim}{\mbox
{{\raisebox{-0.4ex}{$\stackrel{>}{{\scriptstyle\sim}}$}}}}

\slugcomment{Accepted by ApJ Letters}

\shorttitle{Images of an equatorial outflow in SS433}
\shortauthors{Blundell et al.}

\begin{document}

\title{Images of an equatorial outflow in SS433}

\author{Katherine M.\ Blundell\altaffilmark{1},
  Amy J.\ Mioduszewski\altaffilmark{2}, Tom W.\ B.\
Muxlow\altaffilmark{3},\\ Philipp Podsiadlowski\altaffilmark{1} \&
Michael P.\ Rupen\altaffilmark{2}} 
\altaffiltext{1}{University of Oxford, Astrophysics, Keble Road, Oxford,
OX1 3RH, U.K.} 
\altaffiltext{2}{National Radio Astronomy Observatory, Soccoro, NM
87801, U.S.A.} 
\altaffiltext{3}{University of Manchester, MERLIN/VLBI National
  Facility, Jodrell Bank Observatory, Cheshire, SK11 9DL, U.K.}

\begin{abstract}
We have imaged the X-ray binary SS\,433 with unprecedented
Fourier-plane coverage at 6\,cm using simultaneously the VLBA, MERLIN,
and the VLA, and also at 20\,cm with the VLBA.  At both wavelengths we
have securely detected smooth, low-surface brightness emission having
the appearance of a `ruff' or collar attached perpendicularly to the
well-studied knotty jets in this system, extending over at least a few
hundred AU.  We interpret this smooth emission as a wind-like outflow
from the binary, and discuss its implications for the present
evolutionary stage of this system.
\end{abstract}

\keywords{Stars: Binaries: Close, Radio Continuum: Stars, Stars:
Individual: SS433}

\section{Introduction}
\label{sec:intro}

SS\,433 is famous as the first known relativistic jet source in our
Galaxy.  Red- and blue-shifted optical lines, indicating velocities up
to $0.26c$, were discovered by \citet{Mar79a,Mar79b} and interpreted
as gas accelerated by oppositely-directed jets \citep{Fab79,Mil79}.
\citet{Mar84} successfully fit a kinematic precessing jet model,
finding an instrinsic jet speed of $\approx0.26c$, a precession period
of $163\rm\,days$, a cone opening angle of $\sim20^\circ$, and an
inclination to the line-of-sight of $\sim80^\circ$.  This model was
spectacularly confirmed by subsequent radio imaging
\citep[e.g.,][]{Hje81,Ver87,Fej88,Ver93}, which showed precessing twin
jets with structure on scales from milli-arcseconds to arcseconds.
X-ray emission lines were also discovered which mimic the behavior of
the `moving' optical lines \citep{Kot94,Mar01}.  On the basis of these
and other photometric data, SS\,433 is believed to be a binary
consisting of a compact object [a black hole \citep{Zwi89,Fab90} or a
neutron star \citep{Fil88,D'Od91}] and an O- or B-type star
\citep{Mar84}.  Distance estimates range from 3.1\,kpc \citep{Dub98}
to 5.5\,kpc \citep{Hje81}.  Here we report the first results from
simultaneous radio observations with the Very Long Baseline Array
(VLBA), the Multi-Element Radio Linked Interferometer Network
(MERLIN), and the Very Large Array (VLA).  In addition to the
well-known relativistic jets, which will be examined in a series of
companion papers, these data reveal smooth radio emission centered on
the radio core, extending in a huge `ruff' or collar up to $\gtsim\
40$\,milli-arcsec either side of the center.

\section{Observations and Images}

We observed SS\,433 on 6, 7, and 8 March 1998 for four hours each day
with 26 antennas of the VLA; for 10\,hrs each day with the
6-antenna MERLIN array; and for 12\,hrs each day with the 10-antenna
VLBA together with one VLA antenna.  The VLA interleaved observations
at $\sim20$, 6, and 2\,cm; the VLBA, at $\sim20$, 13, and 6\,cm; and
MERLIN observed only at $\sim6\rm\,cm$. The images shown here were
made from the 7~March 1998 data, combining all the arrays at
$\sim6\rm\,cm$, and using the VLBA alone at $\sim20\rm\,cm$.  Images
from the independent data taken on 6 and 8 March 1998
confirm all the features reported here.

The 6\,cm image shown in Figure 1 was made by combining the 7 March
1998 data from all three arrays.  MERLIN observed at 4.993\,GHz with a
bandwidth of 15\,MHz in each of the two independent circular
polarizations, using J\,1907+0127 as a phase-reference source.  The
VLBA observed at four independent frequency bands (IF pairs), each
8\,MHz wide, in each of the two circular polarizations.  Two IF pairs
(centered on 4.990 and 4.998\,GHz) directly overlap
MERLIN's frequency coverage; the other two are not considered here.
J\,1907+0127 and J\,1929+0507 were used as phase-reference sources,
sandwiching 120\,s scans on SS\,433 between 70\,s scans on those two
calibrators.  The total time on SS\,433 (after flagging) at this
wavelength was 2.7\,hrs.  The VLA observed with two IF pairs, each
50\,MHz wide, in each of the two circular polarizations, centered on
4985.1 and 4614.9\,MHz. The calibrator 1950+081 was used for phase
referencing, while occasional observations of 3C\,286 set the
fundamental flux scale.  Initial phase and flux calibration, and
fringe-finding for the VLBA, were carried out using standard
procedures in NRAO's Astronomical Imaging Processing System software.
The (independently derived) flux scales of the three instruments
agreed to better than a few per cent, as determined in overlapping
sections of the {\it uv}-plane.  Finally, the MERLIN data, the
overlapping IF pairs from the VLBA, and the higher-frequency IF pair
from the VLA were concatenated to produce a final, combined data set
with full sensitivity on spatial scales ranging from $\sim10^{-3}$ to
$\sim10$\,arcseconds.  

The VLBA observed SS\,433 using four IF pairs, each covering 8\,MHz in
each of the two independent circular polarizations.  The central
frequencies were 1.404, 1.505, 1.651, and 1.658\,GHz, chosen to be
spread as far apart as possible within the 18--21\,cm band.  The same
observing scheme was used as at 6\,cm.  The image shown
here however is referenced to J\,1928+0507.  To avoid
uncertainties associated with a spatially-variable spectral index,
each IF pair was independently imaged (e.g.\ Fig.\,\ref{fig:lband}).

\psfig{figure=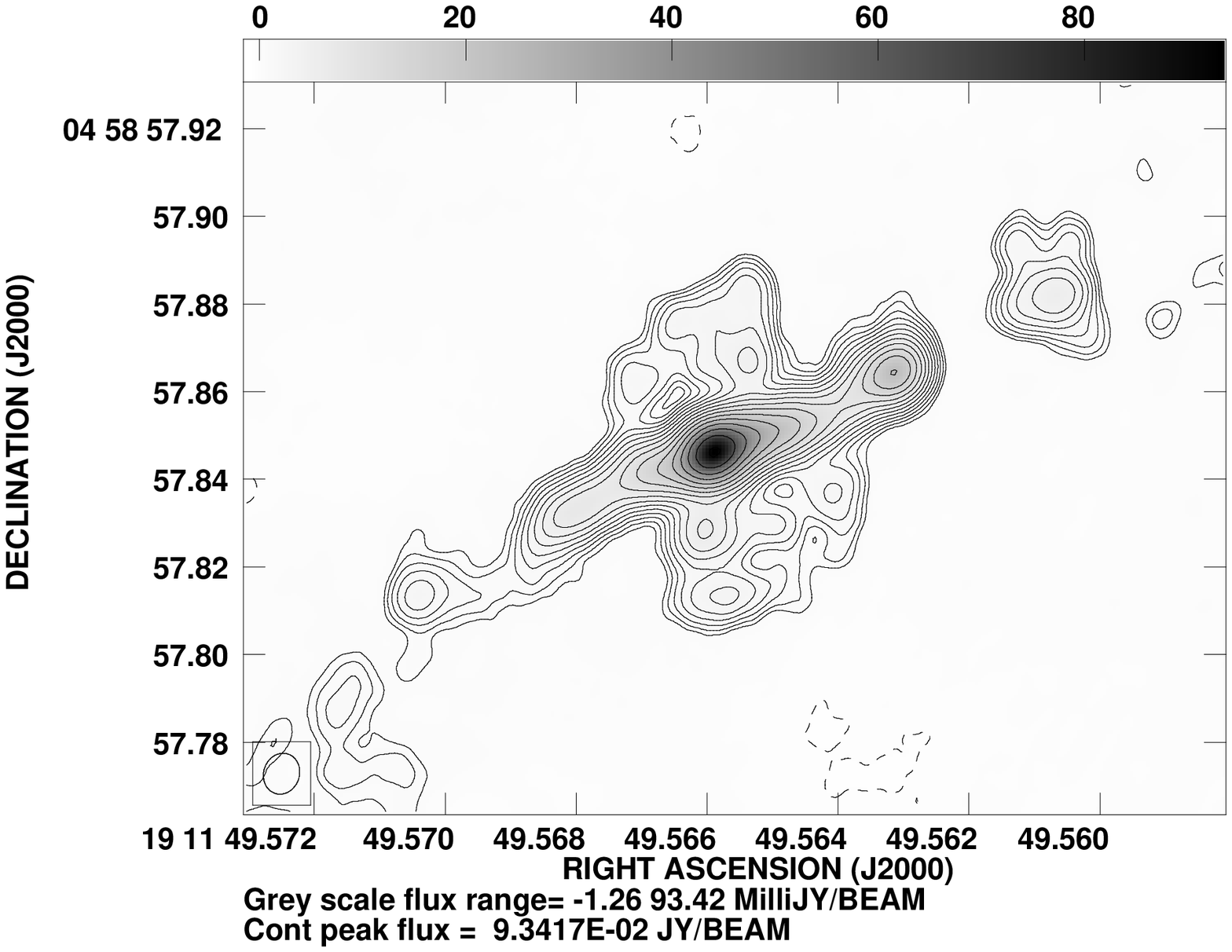,width=8.7cm,angle=0,clip=t} 
\figcaption{4.99\,GHz image of SS\,433 on 7 March 1998, combining
VLBA, MERLIN, and VLA data.  The restoring beam is
$9.5\times8.2\rm\,milliarcseconds$ (FWHM), oriented at a position
angle of $-18\fdg1$.  Contours are $0.5\times\sqrt{2}^n\,\rm
mJy/beam$, $n$= 1, 2, \dots 
\label{fig:cband} 
}

\psfig{figure=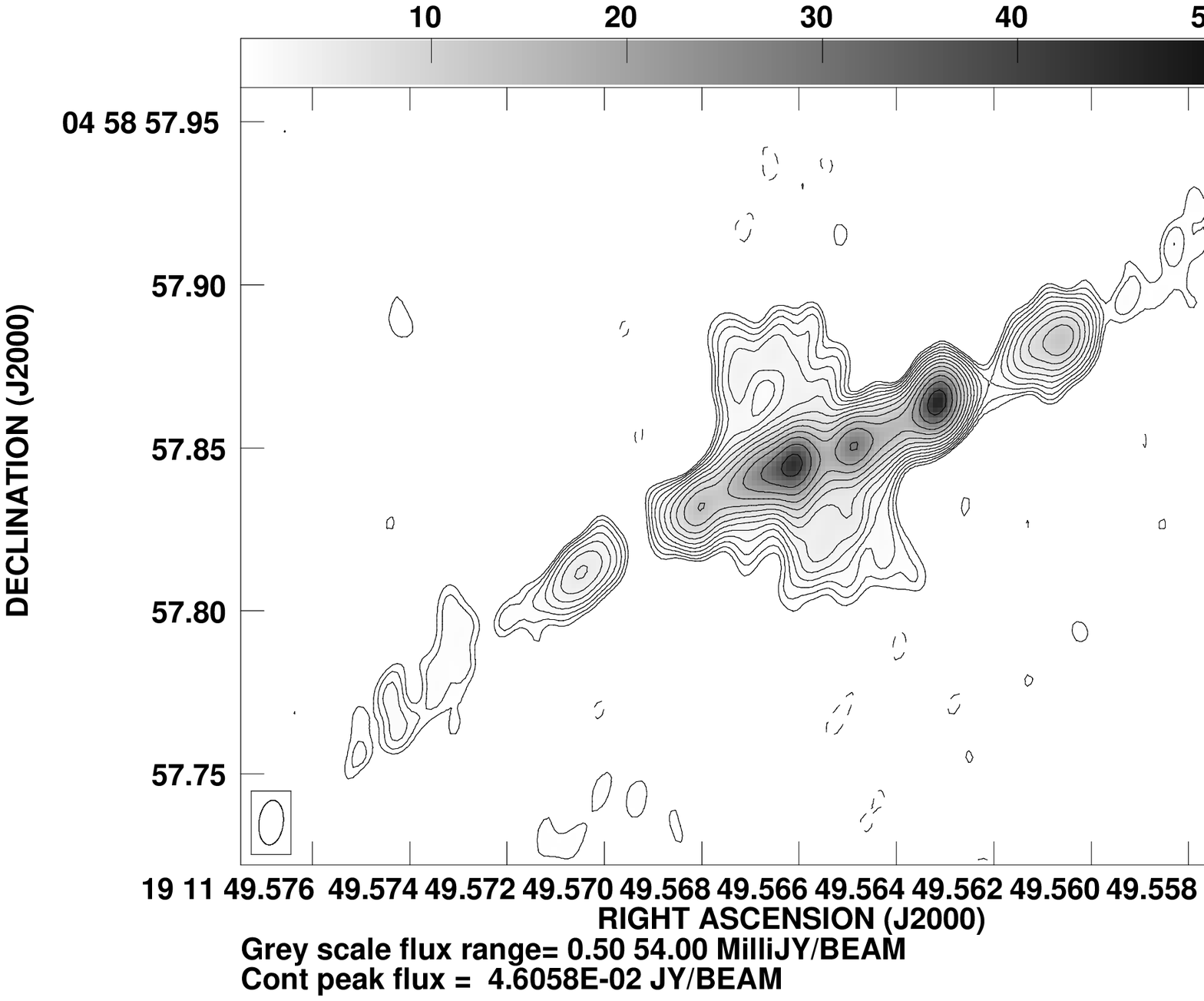,height=6.8cm,angle=0}
\figcaption{1.6\,GHz VLBA image of SS\,433 on 7 March 1998. The
  restoring beam is $13.8\times7.5\rm\,milliarcseconds$ (FWHM),
  oriented at a position angle of $-6\fdg7$.  Contours are
  $0.5\times\sqrt{2}^n\,\rm mJy/beam$, $n$= 1, 2, \dots, 13.
  \label{fig:lband} }

\section{A Thermal `Ruff' About the Center of SS\,433}

Apart from the well-known knotty jet, the most striking feature of
these new images is the smooth emission seen perpendicular to the jet
at its center.  Since this emission appears to encircle the main jet,
we refer to it as a `ruff'.

\subsection{Is the `Ruff' Real?}
Careful testing persuades us that the `ruff' emission is real, and not
some calibration or imaging artifact: self-calibration (avoiding the
ruff emission itself) and careful data editing cannot remove the
emission, which is clearly visible after a single round of phase-only
self-calibration based on the brightest portions of the jet.  The
emission is seen at consistent flux levels, positions, and
orientations: in MERLIN and VLBA data reduced independently, and in
data from three consecutive days reduced separately.  The `ruff'
appears with similar size, brightness asymmetry, position, and flux
density (see below), in images made in the individual IF pairs
covering 18--21\,cm and 6\,cm.

\subsection{Previous Observations}

Paragi and co-workers were the first to report firm evidence for
equatorial emission distinct from the jet itself
\citep{Par98,Par99a,Par99b}, from VLBA data taken in May~1995.  There
are some significant differences in what we see, for example they
detect equatorial emission which appears to be rather more blobby than
smooth; and while our images suggest a `halo' which connects smoothly
to (or around) the jet, their naturally-weighted 1.6\,GHz image
(fig.\,1 in \citet{Par99a}) shows a gap of more than 30\,mas between
the two.  It may be that these differences indicate genuine
variability in this structure.

Our discovery is the first reported detection of such a smooth feature
in SS\,433 extended perpendicular to the jets.  In part this is
because of the high-quality Fourier sampling of these observations,
which gives good sensitivity over a wide-range of spatial scales,
and also provides much higher fidelity images, less vulnerable to
calibration and imaging artefacts.

Evidence of extended emission on much larger angular scales may be
seen approximately north and south of the central circular blob in the
330\,MHz image of \citet[figs 1a \& 1b]{Dub98}.

\subsection{Characterizing the `Ruff' Emission}

In order to determine the spectral shape of the `ruff' emission we
chose {\it uv}-weighting schemes which gave roughly the same dirty
beams for both the 6 and the 18\,cm data, CLEANed the images to
somewhat below the rms noise levels, and finally convolved the
resulting (model+residual) images to a common $10 \times 10$\,mas
Gaussian beam.  The resulting total flux densities, measured in
identical boxes in all images, which were chosen to avoid the jet but
include the full `ruff' emission, are shown in
Figure\,\ref{fig:alpha}$a$.  Note that these are {\it lower limits} to
the total `ruff' emission, since we are excluding any such emission
which overlaps with the jet at this spatial resolution.  The spectral
index for the combined (northern+southern) emission is
$\alpha=-0.12\pm0.02$ ($S_{\nu} \propto \nu^\alpha$, where $S_{\nu}$
is the flux density at frequency $\nu$). This is an interesting result
since most resolved synchrotron sources are characterized by $\alpha <
-0.4$; indeed, $\alpha=-0.1$ is normally considered the signature of
thermal bremsstrahlung emission as is often observed in outflows from
symbiotic binaries \citep{Sea84,Mik01}.  The complication here is that
the peak surface brightness corresponds to a brightness temperature of
$(2-4)\times10^7\,\rm K$ at 18\,cm, implying a similar {\it lower
limit} to the physical temperature of a thermally-emitting plasma.

The distribution of the flux density perpendicular to the jet is shown
in Figure\,\ref{fig:alpha}$b$, which suggests that the spectral index is
indeed almost flat throughout the ruff, and further shows that the
emission extends to \gtsim\ $40\,\rm mas$ at our sensitivity, or
$\sim120\left(d/3\,\rm kpc\right)\,AU$.  Note also that the ruff is 
roughly symmetric about the jet.

\subsection{Implications for X-ray emission}

If the radio emission is bremsstrahlung from a thermal population of
particles then that same population should co-spatially emit X-rays.
The X-ray luminosity is given by $L_{\rm x} = L_{\rm rad} \times \exp
({-h\nu_{\rm x}/kT})$ where $T$ is the temperature of the particles
(approximated to be the radio brightness temperature), $k$ and $h$ are
the Boltzmann and Planck constants, $\nu_{\rm x}$ is the
lower frequency of the X-rays whose luminosity $L_{\rm x}$ is
predicted from the radio luminosity $L_{\rm rad}$.  We measure a
brightness temperature $T_{\rm B} \gtsim 10^7\,{\rm K}$ which
over-predicts the X-ray luminosity compared with that observed
\citep{Mar84}.  Allowing for a significant fraction of the observed
X-rays being emitted by the jets means the discrepancy could be as
much as an order of magnitude.  This could in principle be due to the
presence of neutral material in the vicinity of the ruff which would
absorb X-rays but not radio emission.  Alternatively, it may suggest
that the particle population is not Maxwellian.

\section{The origin of the smooth emission}

\subsection{Theoretical background}
\label{sec:outflow}
\def\Ms{\hbox{$\,M_{\odot}$}}
\def\yr{\hbox{$\,\hbox{yr}$}}

The most straightforward interpretation of the radio emission is that
it arises from mass outflow from the binary system that is enhanced
towards the orbital plane. Such mass loss could either (i) come from
the companion (most likely an O or B star), (ii) be a disk wind from
the outer parts of the accretion disk or (iii) arise from mass loss
from a proto-common envelope surrounding the binary components. The
detection of this mass loss may have rather important implications for
our understanding of the evolutionary state of this unique system. It
has been a long-standing puzzle how SS\,433 can survive so long in a
phase of extreme mass transfer ($\dot{M} \ga 10^{-5}\Ms\yr^{-1}$)
without entering into a common envelope phase where the compact object
spirals completely into the massive companion (for a recent discussion
see King, Taam \& Begelman 2000).  Since the theoretically predicted
mass-transfer rate exceeds even the estimated mass-loss rate in the
jets ($\dot{M}\sim 10^{-6}\Ms\yr^{-1}$; Begelman et al.\ 1980), King
et al.\ (2000) proposed that most of this transferred mass is lost
from the system in a radiation-pressure driven wind from the outer
parts of the accretion disk \citep[see also][]{King99}.  A related
problem exists in some intermediate-mass X-ray binaries (IMXBs).
Models of the IMXB Cyg X-2 \citep{King99a,Pod00,Kolb00,Taur00} show
that the system must have passed through a phase where the
mass-transfer rate was $\sim 10^{-5}\Ms\yr^{-1}$, exceeding the
Eddington luminosity of the accreting star by many orders of
magnitude, without entering into a common-envelope phase, and where
almost all the transferred mass must have been lost from the
system. The observed emission in SS\,433 presented here may provide
direct evidence of how such mass loss takes place.

\subsection{Evidence for mass out-flows}

The existence of a disk-like outflow was first postulated by
\citet{Zwi91} to explain the variation with precession phase of the
secondary minimum in the photometric light curve.  \citet{Fab93}
proposed a disk-like expanding envelope caused by mass-loss from the
outer Lagrangian point L2 to explain the blue-shifted absorption lines
of H\,I, He\,I and Fe\,II (see also \citet{Mam80}, whose spectrum
shows that all the emission lines seen in SS\,433 have P-Cygni
profiles indicating the presence of outflowing gas).  \citet{Fil88}
observe a remarkable double peaked structure for the Paschen lines,
with speeds close to 300\,${\rm km\,s^{-1}}$.

\subsection{Estimates of $\dot{M}$ and the windspeed}
\label{sec:stellarwind}
If we assume that the observed radio emission is due to
bremsstrahlung, we can obtain a rough estimate for the mass-loss rate,
$\dot{M}$, in this equatorial outflow. For this purpose, we assume
that the outflow is radial but confined to an angle $\alpha$ with
respect to the orbital plane of the binary. For a simple wind
mass-loss law, the mass density, $\rho$, of the outflow then depends
on the distance $r$ from the system according to
$\rho=\dot{M}/(4\pi\,\sin\alpha\,r^2\,v_{\infty})$, where $v_{\infty}$
is the outflow velocity at infinity. At a particular radio frequency
$\nu$, the outflow will be optically thick to some distance
$\bar{r}_{\nu}$.

Assuming that we see all the radio emission from the optically thin
part of the outflow and are observing the system close to the orbital
plane (both assumptions are only approximately true and ignore
geometrical complications), a rough estimate for the
mass-outflow rate is
\begin{eqnarray} 
\dot{M}&\simeq& 1.6\times 10^{-4}\,M_{\odot}\,\mbox{yr}^{-1}\,\,
\\
&&\hspace{1cm}\times S_{50}^{3/4}\,d_{3}^{3/2}\,v_{300}\,\nu_{1.4}^{-1/2}\,
\bar{g}_{10}^{-1/2}\, (\sin\alpha)_{30}^{1/4},\nonumber 
\end{eqnarray}  
where $S_{50} = S_\nu/50$\,mJy, $d_{3}=d/3\,$kpc, $v_{300}=
v_\infty/300\,$km\,s$^{-1}$, $\nu_{1.4}=\nu/1.4\,$GHz, $\bar{g}_{10}=
\bar{g}/10$ ($\bar{g}$ is the Gaunt factor for free-free emission;
see e.g., Rybicki \& Lightman 1979),
$(\sin\alpha)_{30}=\sin\alpha/\sin 30^\circ$.

One of the major uncertainties in this estimate is the velocity of the
outflow, though a velocity of $\sim 300\,$km\,s$^{-1}$ is similar to
that of the lines by Filippenko et al.\ (1988) and is close to the
characteristic orbital velocity of SS 433, as one might expect for an
outflow from the binary system rather than either binary component.
Furthermore, if this outflow started soon after the supernova
explosion which formed the compact object $\sim 10^4\,$yr ago and
whose impressively circular remnant is seen clearly in the images of
Dubner et al.\ (1988), a velocity of $\sim 300\,$km\,s$^{-1}$ implies
an extent of the outflow of $\sim 3\,$arcmin (for $d=3\,$kpc). Indeed,
this is exactly the size of the extended smooth emission seen by
Dubner et al.\ (1998) and suggests that this may be the outer extent
of the same outflow.
	
The above estimate for $\dot{M}$ would imply that the outflow is
optically thick at a frequency of $\sim 1\,$GHz to a distance of
$r\sim 10^{15}\,$cm and that the radio emission from the central
region of the jet would be somewhat attenuated (although this will
also depend on the exact geometry of the outflow and its orientation
to the line of sight).

It also suggests that the outflow is moderately optically thick in the
optical and that part of the observed visual extinction to the system
($A_V = 7.8$, Margon 1984) may be due to the outflow, as 
postulated by Zwitter et al.\ (1991).

The inferred mass-loss rate, $\dot{M}\sim
10^{-4}\,M_{\odot}\,$yr$^{-1}$, is much higher than any reasonable
mass-loss rate from an O-star primary and suggests that it is
connected with the unusual short-lived phase SS\,433 is experiencing
(see \S~4.1). It could be mass loss from a common envelope that has
already started to form around the binary, or a hot coronal wind from
the outer parts of the accretion disk driven, e.g., by the
X-ray irradiation from the central compact source.

\section{Summary}

With unique sampling in the UV-plane, we have imaged the SS\,433
system at 6\,cm and 20\,cm and securely detected at both wavelengths
smooth emission extending over a few hundred AU perpendicular to the
jet axis.  The most likely interpretation of this radiation is
emission from matter which has been ejected from the disk as a thermal
wind with an outward speed of $\sim 300\,{\rm km\,s^{-1}}$.

\acknowledgments

K.M.B.\ thanks the Royal Society for a University Research Fellowship.
MERLIN is a U.K.\ national facility operated by the University of
Manchester on behalf of PPARC.  The VLBA and VLA are facilities of
NRAO operated by AUI, under cooperative agreement with the NSF.

\centerline{
\psfig{figure=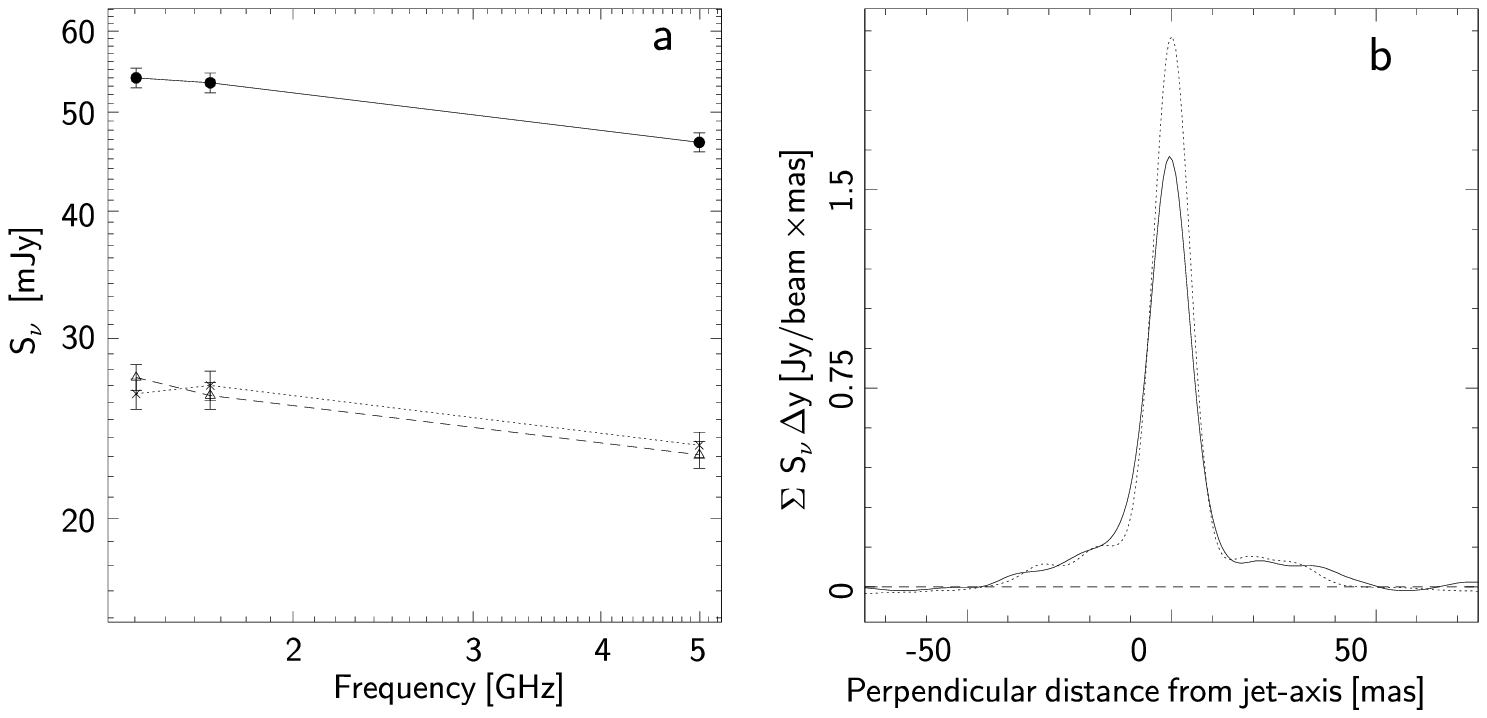,height=12cm,angle=0,clip=t}
}
\figcaption{{\em a:} The total flux density
in the `ruff' emission as a function of frequency (see text).
Crosses: northern emission; triangles: southern emission; filled
circles: sum of northern and southern emission.  {\em b:} The flux
density integrated over 40\,mas strips parallel to the jet, as a
function of distance perpendicular to the jet, at 10\,mas resolution.
The solid line is 18\,cm, the dotted line the 6\,cm data.  Note the
flat spectrum of the `ruff' emission, compared to the inverted
spectrum of the (self-absorbed) jet core.
\label{fig:alpha}
} 

\end{document}